\documentclass[prd,aps,onecolumn,showpacs,preprintnumbers,amsmath,amssymb,superscriptaddress,nofootinbib]{revtex4-1}
\usepackage[dvips]{graphicx}
\usepackage{epsf}
\usepackage{amsmath}
\usepackage{amssymb}
\usepackage{amsfonts}
\usepackage{times}
\usepackage[usenames]{color}
\usepackage[normalem]{ulem}
\usepackage[T1]{fontenc}
\usepackage[dvipsnames]{xcolor}

\usepackage{dcolumn}% Align table columns on decimal point
\usepackage{bm}% bold math
\usepackage{color}

\begin{document}

\title{Weighing neutrinos in $f(R)$ gravity in light of BICEP2}

\author{Xiao-ying Zhou}
\email[Email address: ]{zhouxiaoying@fudan.edu.cn}
\affiliation{Department of Physics, Fudan University, 200433, Shanghai, P. R. China}

\author{Jian-hua He}
\email[Email address: ]{jianhua.he@brera.inaf.it}
\affiliation{INAF-Observatorio Astronomico, di Brera, Via Emilio Bianchi, 46, I-23807, Merate (LC), Italy}


\begin{abstract}
We constrain the neutrino mass in $f(R)$ gravity using the latest observations from the Planck, BAO and BICEP2 data. We find that the measurement on the B-modes can break the degeneracy between the massive neutrinos and the $f(R)$ gravity. We find a non-zero value of the Compton wavelengths $B_{0}$ at a $68\%$ confidence level for the $f(R)$ model in the presence of massive neutrinos when the BICEP2 data is used. Furthermore, the tension on the tensor-to-scalar ratios between the measured values from Plank and BICEP2 is significantly reconciled in our model.
\end{abstract}

\maketitle
\section{Introduction}
The detected neutrino oscillations in atmospheric, solar and reactor neutrino experiments\cite{experiment} implies that neutrino has a non-zero absolute mass. If the neutrino does have the absolute mass, it will be the lowest energy particle in the extensions of the Standard Model of particle physics. The determination of the neutrino mass therefore is one of the greatest unsolved problems in modern physics. However, the neutrino experiments are able to place a lower limit on the effective neutrino mass, which depends on the hierarchy of the neutrino mass spectra\cite{Hitoshi}(also see\cite{nu_review} for reviews). The laboratory experiments cannot exactly pin down the absolute mass scale of neutrinos.

On the other hand, cosmological constraints on the neutrino mass are highly complementary to
particle physics. Cosmology provides a unique laboratory for studying neutrino mass. The neutrino mass has already been constrained up to an unprecedented accuracy simply by the cosmological probes~\cite{neutrinolcdm}. It is even more promising that the allowed neutrino mass window could be closed by forthcoming cosmological surveys~\cite{luca}. Despite the notable success in this effort, the cosmological neutrino constraints are highly model dependent. The current constraints are usually obtained in the context of a $\Lambda$CDM model or a dynamical dark energy model~\cite{neutrinolcdm}. It is important to test the robustness of such constraints against assumptions about the nature of gravity and dark energy. The observed cosmic acceleration may also be explained without dark energy if our theory of gravity is modified. It is therefore reasonable to study cosmological neutrino constraints in the framework of modified gravity. The impact of massive neutrinos on the linear growth history in the $f(R)$ gravity has been studied in the literature\cite{neutrinofr,gbz,Henu,JD}. Strong degeneracy has been found between the late time growth of the $f(R)$ gravity and the massive neutrinos\cite{neutrinofr,gbz,Henu,JD}. Such degeneracy has also been found at the non-linear level based on N-body simulations~\cite{baldi}. Thus, tighter constraints on the neutrino mass can only be achieved if there is an independent constraint on the neutrino mass or the modified gravity models are well constrained by the local tests of gravity.

Very recently, BICEP2 group have announced a robust detection on the $B-$mode power spectrum in the cosmic microwave background(CMB) at a significance of $>5\sigma$~\cite{BICEP2}. The finding provides the long-sought evidence of inflation and the first detection of gravitational waves' action on matter. The detection of gravitational waves also provide an unique opportunity to break the degeneracy between the massive neutrino and modified gravity. The reason for this is that, as we shall show later in this paper, the $f(R)$ gravity has insignificant impact on the evolution of the tensor perturbation. However, the massive neutrinos contribute directly to the anisotropic stress and has direct impact on the angular power spectra of the B-mode. The measurement of the B-modes thus can break the degeneracy between the massive neutrinos and modified gravity. The aim of this paper is therefore to test the robustness of the measurement  of the B-modes from BICEP2 in breaking such degeneracy. We will discuss the tensor perturbation in the $f(R)$ gravity first and then constrain the neutrino mass using the observations from BICEP2~\cite{BICEP2}, Planck~\cite{planck} and Baryon Acoustic Oscillations (BAO) surveys.

This paper is organized as follows: in Sec.~\ref{frtensor}, we will briefly introduce our $f(R)$ model and review the tensor perturbation in the $f(R)$ gravity. In Sec.~\ref{Data}, we will introduce the observational data used in our analysis. In Sec.~\ref{Num}, we will
present the fitting results for our $f(R)$ model. In Sec.~\ref{con}, we will summarize and conclude this work.

\section{tensor perturbation in $f(R)$ gravity\label{frtensor}}
We consider the action as
\begin{equation}
S=\frac{1}{2\kappa^2}\int d^4x\sqrt{-g}[R+f(R)]+\int
d^4x\mathcal{L}^{(m)}\label{action}\quad,
\end{equation}
where $\kappa^2=8\pi G$ with $G$ being Newton's constant. $g$ is the determinant of the metric $g_{\mu\nu}$. $\mathcal{L}^{(m)}$ is the Lagrangian density for the matter fields. $f(R)$ is an arbitrary function of the Ricci scalar curvature $R$~\cite{fr1,fr2,fr3,fr4,fr5,fr6,fr7,fr8,fr9,fr10,fr11,fr12}. For the background cosmological evolution, we consider the Friedmann-Robertson-Walker metric
\begin{equation}
ds^2=a^2[-d\tau^2+d\sigma^2]\quad,
\end{equation}
where $d\sigma^2$ is the conformal space-like hypersurface with a constant curvature $R^{(3)}=6K$
\begin{equation}
d\sigma^2=\frac{dr^2}{1-Kr^2}+r^2(d\theta^2+sin^2\theta d\phi^2)\quad .
\end{equation}
The modified equation of motion for the Universe is given by\cite{frreview}
\begin{equation}
\ddot{F}+2F\dot{H}-H\dot{F}-\frac{2K}{a^2}F=-\kappa^2(\rho+p)\quad.\label{dfield}
\end{equation}
where $F=\frac{df(R)}{dR}$, the dot denotes the time derivative with respect to the cosmic time $t$ and $\rho $ is the total energy density of the matter fields. $p$ is the total pressure in the Universe. In this work, we study a specific family of $f(R)$ models that have the same background as the $\Lambda$CDM model\cite{frmodel}. Equation (\ref{dfield}) can be solved numerically given the initial condition at the deep matter dominated epoch\cite{frlinear}
\begin{equation}
F(x)\sim1+D(e^{3x})^{p_+}\quad,\frac{dF(x)}{dx}\sim3Dp_+(e^{3x})^{p_+}\quad,\label{initial}
\end{equation}
where $x=\ln a$ and the index is defined by $p_+=\frac{5+\sqrt{73}}{12}$. $D$ is a free parameter that characterizes the $f(R)$ model. This family of $f(R)$ models
can also be equivalently characterized by the Compton wavelengths in units of the Hubble scale~\cite{Song}
\begin{equation}
B_0=\frac{f_{RR}}{F}\frac{dR}{dx}\frac{H}{\frac{dH}{dx}}(a=1)\quad.
\end{equation}
We choose the initial time for Eq.(\ref{initial}) at $a_i=0.02$.

For the perturbed spacetime, we consider both the scalar perturbation and the tensor perturbation. The scalar perturbation within the framework of $f(R)$ gravity has been well studied in Ref.~\cite{frlinear} as well as Refs.~\cite{frperturbationreview,bean} . Readers are referred to these papers for details. For simplicity, we will not present and repeat here. In this work, we will only focus on the tensor perturbation.

The perturbed line element for the tensor perturbation is given by
\begin{equation}
ds^2=-a(\tau)^2d\tau^2+a(\tau)^2(\gamma_{ij}+2H^{T}_{ij})dx^idx^j\quad,
\end{equation}
where $\tau$ is the conformal time. $H^{T}_{ij}$ is a traceless($H^{Ti}_{i}=0$), divergencefree ($\nabla^{i}H^{T}_{ij}=0$), symmetric ($H^{T}_{ij}=H^{T}_{ji}$) tensor field. In the fourier space, the perturbed modified Einstein equation yields the modified wave equation for gravitational waves~\cite{Hwang}
\begin{equation}
H_T{''}+(2\mathcal{H}+\frac{F'}{F})H_T'+(k^2+2K)H_T=\frac{\kappa^2a^2p\pi^{(2)}}{F}\quad,\label{tensor}
\end{equation}
where $F$ is the background scalar field and the prime denotes the derivative with respect to the conformal time $\tau$. $\pi^{(2)}$ is the anisotropic stress. From Eq.(\ref{tensor}), we can see that the $f(R)$ gravity affects the tensor perturbation $H_T$ only through the background field $F$. At early times, this effect can be neglected because the viable $f(R)$ models should go back to the $\Lambda$CDM model at early times $\lim_{R\rightarrow} +\infty F\rightarrow 1$. On the other hand, the massive neutrinos contribute directly to the anisotropic stress $\pi^{(2)}$, substantially changing the evolution of $H_T$. The massive neutrinos has significant impact on the B-mode of the CMB angular power spectra. The degeneracy between the $f(R)$ gravity and massive neutrinos hence no longer exists in the tensor perturbations.
The measurement of the B-mode therefore is a promising way to constrain the massive neutrinos in the $f(R)$ gravity. In practice, we solve Eq.(\ref{tensor}) numerically basing on our \texttt{FRcamb} code~\cite{frlinear}~\cite{FRcamb}. We choose the initial time for the scalar field $F$ starting to enter Eq.(\ref{tensor}) at $a_{ini}=0.02$, before which we set $F=1$ and $F'=0$. Along with our previous work studying the linear scalar perturbations in the $f(R)$ gravity~\cite{frlinear}, our current work completes the tensor perturbation in our \texttt{FRcamb} code~\cite{FRcamb}.

\section{Current Observational Data\label{Data}}
In this work, we use three probes from the measurements of the CMB, including the Planck\cite{planck}, BICEP2\cite{BICEP2}, and the high-$l$ data from the Atacama Cosmology Telescope(ACT)\cite{ACT} and the South Pole Telescope(SPT)\cite{SPT}. For the Planck data, we only use the data on the temperature angular power spectra. We do not include the CMB lensing data from the Planck in our analysis. For the BICEPE2 data, we adopt 9 bins of E and B polarization data in the range of $30<l<150$. In our analysis, we also use the WMAP nine-year polarization data~\cite{wmap9} along with the Planck temperature data.

Since the family of $f(R)$ model studied in this work does not change the scales of the BAO peak in the real spacetime~\cite{Henu,frlinear}, in addition to the CMB data, we also use the measurements from the BAO surveys. The BAO surveys measure the distance ratio between $r_s(z_{\rm drag})$ and $D_v(z)$,
where $r_s(z_{\rm drag})$ is the comoving sound horizon at the baryon drag epoch
and $D_{v}(z)$ is a combination of the angular-diameter distance $D_{A}(z)$ and the Hubble parameter $H(z)$.
\begin{equation}
D_{v}(z)=\left[(1+z)^2D_{A}^2(z)\frac{cz}{H(z)}\right]^{1/3}\quad.
\end{equation}

We adopt the BAO measurements from four different redshift surveys, following the analysis by the Planck team\cite{planck}:the BOSS DR9 measurement \cite{BAO1} at $z=0.57$;
the 6dF Galaxy Survey measurement \cite{BAO2} at $z=0.1$ ;
the WiggleZ measurement\cite{BAO3} at $z=0.44,0.60$ and $0.73$;
the SDSS DR7 measurement\cite{BAO4} at $z=0.2$ and $z=0.35$ .
\section{Numerical results\label{Num}}
We implement the Markov Chain Monte Carlo analysis on the parameter space for our model, basing on the public available code COSMOMC \cite{mcmc} as well as our \texttt{FRcamb} code~\cite{frlinear}~\cite{FRcamb} which is a modified version of the CAMB code~\cite{CAMB}. Our code now has been updated to include the tensor perturbation in the $f(R)$ gravity. The parameter space of our model is
\begin{equation}
P=(\Omega_bh^2,\Omega_ch^2,100\theta_{\rm MC},\ln[10^{10}A_s],n_s,\tau,\sum m_{\nu},D,r)\quad,
\end{equation}
where $\Omega_bh^2$ and $\Omega_c h^2$ are the physical baryon and cold dark matter energy densities respectively, $A_s$ is the amplitude of the primordial curvature perturbation, $100\theta_{\rm MC}$ is the angular size of the acoustic horizon, $n_s$ is the scalar spectrum power-law index, $\tau$ is the optical depth due to reionization, $\sum m_{\nu}$ is the sum of neutrino mass in eV. $r$ is the tensor to scalar ratio. The pivot scale is set at $k_{s0}=0.05{\rm Mpc^{-1}}$. In this work, we only focus on the total mass of active neutrinos since the family of $f(R)$ models studied in this work has less impact on the effective number of neutrino-like relativistic degrees of freedom $N_{\rm eff}$~\cite{Henu}. $N_{\rm eff}$ is of less interest in our analysis. We set $N_{\rm eff}=3.046$ throughout this work. Further, we do not treat the tensor spectrum power-law index $n_t$ as a free parameter. We adopt the inflation consistency relation $n_t=-A_t/(8A_s)$ in this work. We will sample the parameter $D$ directly and treat $B_0$ as a derived parameter. The priors for the cosmological parameters are listed in table~\ref{priors}.
\begin{table}
\caption{Uniform priors for the cosmological parameters \label{priors}}
\begin{tabular}{c}
\hline
$0.005<\Omega_bh^2<0.1$ \\
$0.001<\Omega_ch^2<0.99$\\
$0.5<100\theta_{\rm MC}<10.0$ \\
$0.01<\tau<0.8$  \\
$0.9<n_s<1.1$  \\
$2.7<\rm{ln}[10^{10}As]<4.0$ \\
$-1.2<D<0$ \\
$0<\sum m_{\nu}<5$ \\
$0<r<1$ \\
\hline
\end{tabular}
\end{table}

\begin{figure}
\includegraphics[width=3in,height=2.8in]{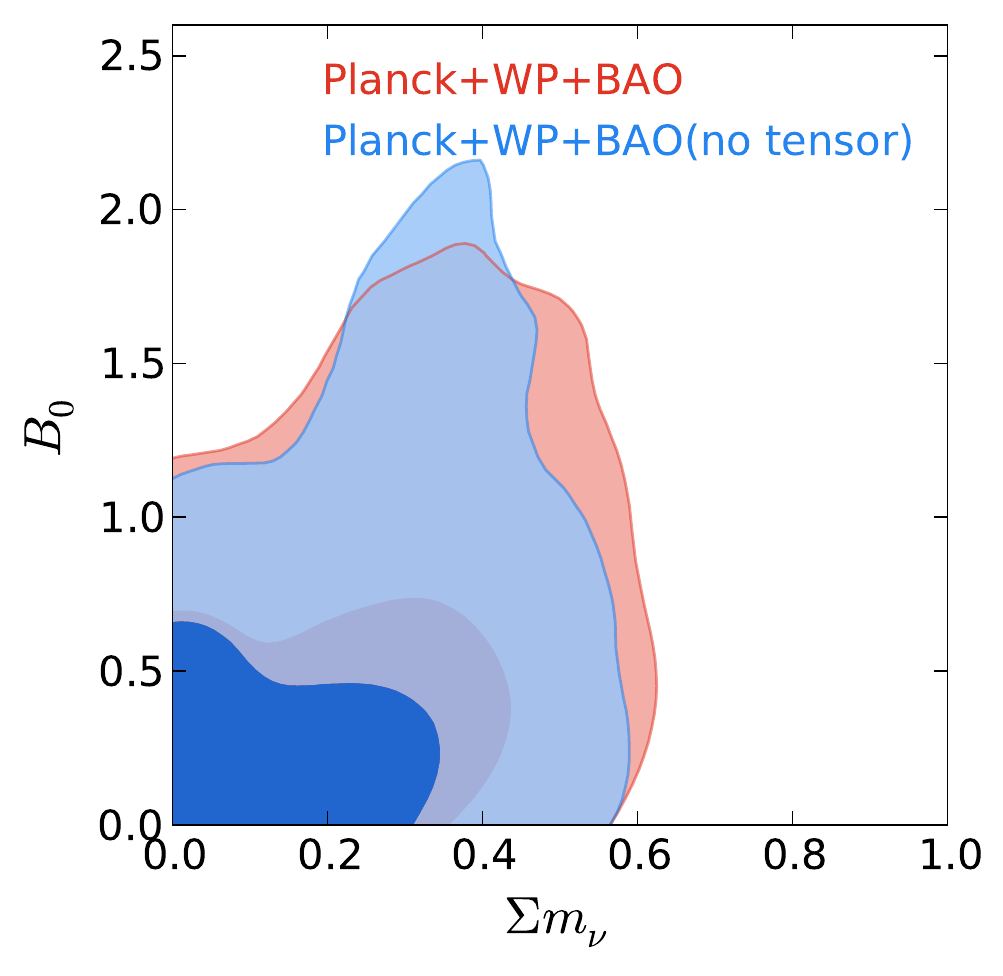}
\includegraphics[width=3in,height=2.8in]{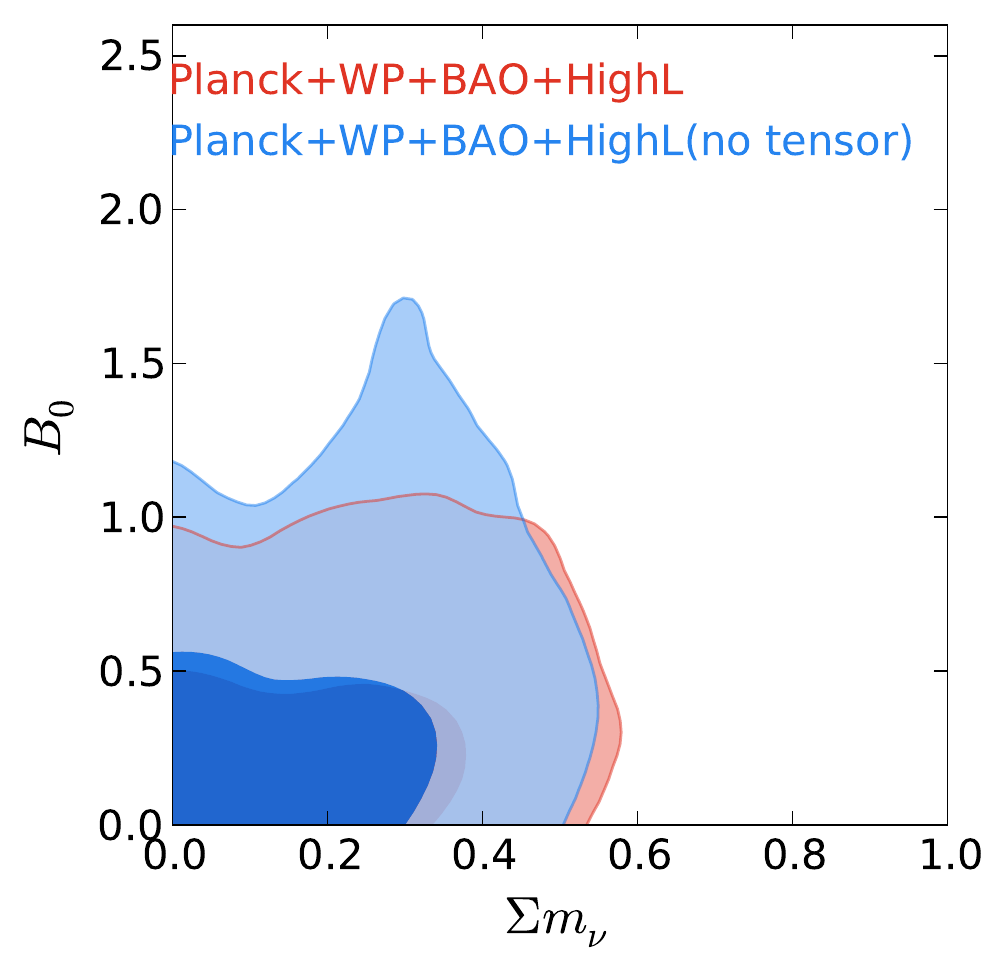}
\caption{Marginalized two-dimensional likelihood ($1, 2\sigma$ contours) constraints on $B_0$ and $\sum m_{\nu}$. In the absent of the tensor perturbation, it is clear that the degeneracy between $B_0$ and $\sum m_{\nu}$ sharpens if $B_0\gtrsim 1 $, as shown in the contour with the data sets of Planck+WP+BAO combination.
}\label{figone}
\end{figure}

First, we shall reinvestigate the degeneracy between the parameters $B_0$ and the total neutrino mass $\sum m_{\nu}$ in the $f(R)$ gravity. For this purpose, we constrain our model without considering the tensor perturbation in the first place. We use the data sets of Planck+WP+BAO combination. We present the fitting results in Table~\ref{fittingresults}. In Fig.\ref{figone}, we show the contour of the marginalized two-dimensional likelihood for $B_0$ and $\sum m_{\nu}$.
The fitting results and the shape of the contour are slightly different from our previous work~\cite{Henu} due to the slight change in the initial condition for the starting time of the $f(R)$ gravity from $a_{ini}=0.03$ to $a_{ini}=0.02$ in this work. As discussed in Ref.~\cite{Henu}, the degeneracy between $B_0$ and $\sum m_{\nu}$ sharpens, if $B_0\gtrsim 1$, due to the compensation of the impact on the ISW effect between the massive neutrinos and the $f(R)$ gravity. Clearly, in Fig.~\ref{figone} there is a long tail in the contour as $B_0>1$. The degeneracy thus is clearly shown in the contour. The poor constraints on the massive neutrinos in the $f(R)$ gravity relative to the case in the $\Lambda$CDM model even using the same date sets are due to such degeneracy. Next, in order to get tighter constraints on the massive neutrinos, we add the HighL data. The fitting results are listed in Table~\ref{fittingresults}. The data sets of Planck+WP+BAO+HighL combination yields the tightest constrains on the total mass of neutrinos in this work
$$\sum m_{\nu}<0.43{\rm eV}(95\%{\rm C.L.};{\rm Planck+WP+BAO+highL})\quad.$$
The result is slightly different from our previous work~\cite{Henu} in which we have obtained $\sum m_{\nu}<0.46{\rm eV}$. The difference is due to the slight change in the initial conditions. However, it has only minor effect on the fitting results. The result is still larger by a factor of two than that obtained in the $\Lambda$CDM case
$$\sum m_{\nu}<0.23{\rm eV}(95\%{\rm C.L.};{\rm Planck+WP+BAO+highL})\quad,$$
as reported by the Planck team~\cite{planck}.

Next, we investigate the impact of the tensor perturbation on the constraints of the massive neutrinos in the $f(R)$ gravity. For comparison, we use the same data sets  and combinations as used in the previous analysis. The fitting results are shown in Table~\ref{fittingresults}. In Fig.1, we overplot the contours with the tensor perturbation for $B_0$ and $\sum m_{\nu}$. We can see that the tensor perturbation slightly looses the constraints on both $B_0$ and $\sum m_{\nu}$ in the case without the HighL data. The degeneracy is alleviated by including the tensor perturbation in the computation.
\begin{table*}
\caption{Impact of the tensor perturbation on the constrains of $\sum m_{\nu}$ and $B_0$. }\label{fittingresults}
\begin{tabular}{c|c|c|c|c}
\hline
\hline
Parameters & Planck+WP+BAO(no tensor) & Planck+WP+BAO & Planck+WP+BAO+HighL(no tensor) &Planck+WP+BAO+HighL\\
\hline
$\sum m_{\nu}[{\rm eV}]$ & $ <0.47$(95\%{\rm C.L.}) & $ <0.51$(95\%{\rm C.L.})&$ <0.43$(95\%{\rm C.L.}) &$ <0.46$(95\%{\rm C.L.})\\
$B_0$ & $ <1.45$(95\%{\rm C.L.}) & $ <1.51$(95\%{\rm C.L.})&$ <1.17$(95\%{\rm C.L.}) &$ <0.91$(95\%{\rm C.L.})\\
\hline
\end{tabular}
\end{table*}

Now we turn to investigate the effect of the BICEP2 data on the fitting results. The BICEP2 data has significant measurement on the B-modes and is expected to be able to break the degeneracy between the massive neutrinos and the $f(R)$ gravity.
The numerical results  are shown in Table~\ref{BICEP}. In Fig.~\ref{figone}, we show the contours of the marginalized two-dimensional likelihood for $B_0$ and $\sum m_{\nu}$ with the BICEP2 data and also compare them with the results that are obtained without using the BICEP2 data. From Fig.~\ref{figone}, we can see that when adding the BICEP2 data, the size and shape of the 1$\sigma$ and 2$\sigma$ contours are significantly changed, compared with these contours without using the BICEP2 data.
It is very interesting to find that  contrary to intuition, when adding the BICEP2 data, instead of tightening the constraints, we find looser constraints on both the massive neutrinos and the $f(R)$ gravity at the $95\%$ confidence level.
\begin{equation}
\left.
\begin{array}{c}
B_0<1.52 \\
\sum m_{\nu}< 0.49{\rm eV}
\end{array}
\right\}
\quad(95\%{\rm C.L.};{\rm Planck+WP+BAO+HighL+BICEP2})\quad.\nonumber
\end{equation}
 The BICEP2 data apparently does not tend to break the degeneracy between $B_0$ and $\sum m_{\nu}$.

\begin{table*}
\caption{Constraints on the total mass of active neutrinos $\sum m_{\nu}$ and $B_0$. }\label{BICEP}
\begin{tabular}{c|c|c|c|c}
\hline
\hline
Parameters & Planck+WP+BAO & Planck+WP+BAO+BICEP2 & Planck+WP+BAO+HighL &Planck+WP+BAO+HighL+BICEP2\\
\hline
$\sum m_{\nu}[{\rm eV}]$ & $ <0.51$(95\%{\rm C.L.}) & $ <0.53$(95\%{\rm C.L.})&$ <0.46$(95\%{\rm C.L.}) &$ <0.49$(95\%{\rm C.L.})\\
$B_0$ & $ <1.51$(95\%{\rm C.L.}) & $ <1.68$(95\%{\rm C.L.})&$ <0.91$(95\%{\rm C.L.}) &$ <1.52$(95\%{\rm C.L.})\\
\hline
\end{tabular}
\end{table*}

\begin{figure}
\includegraphics[width=3in,height=2.8in]{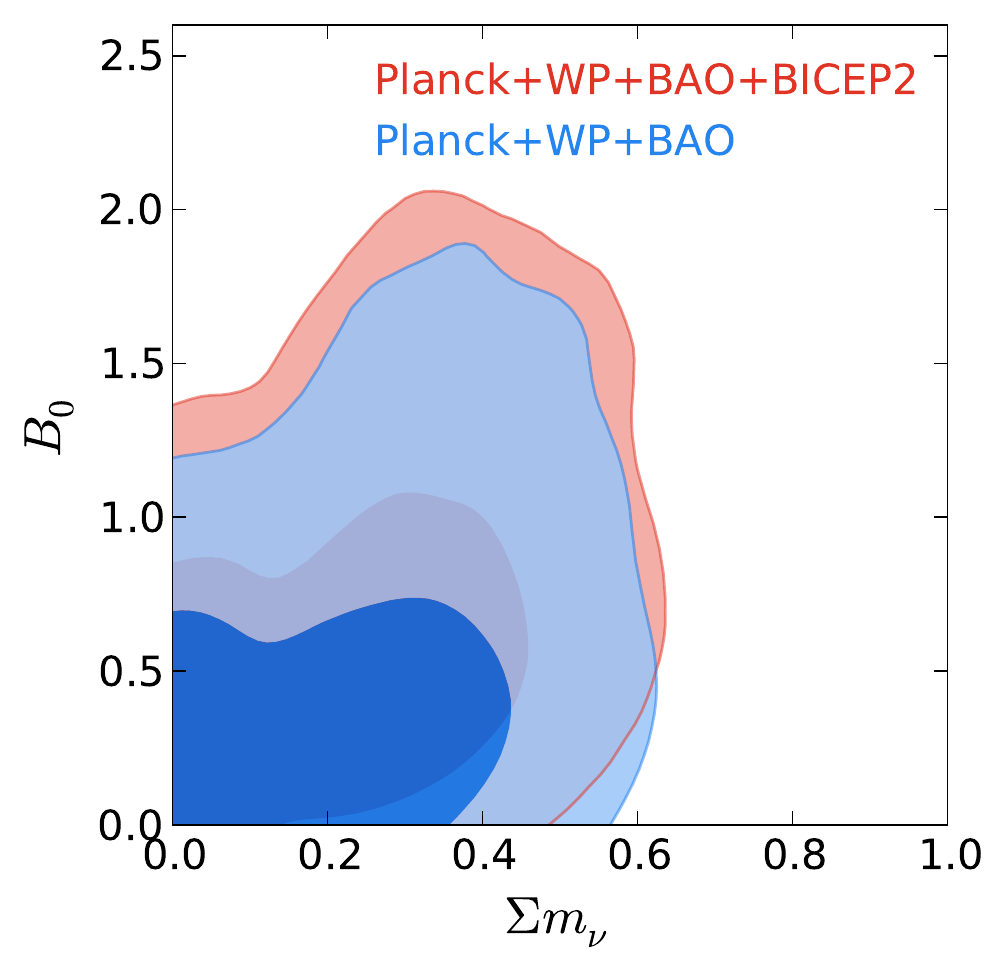}
\includegraphics[width=3in,height=2.8in]{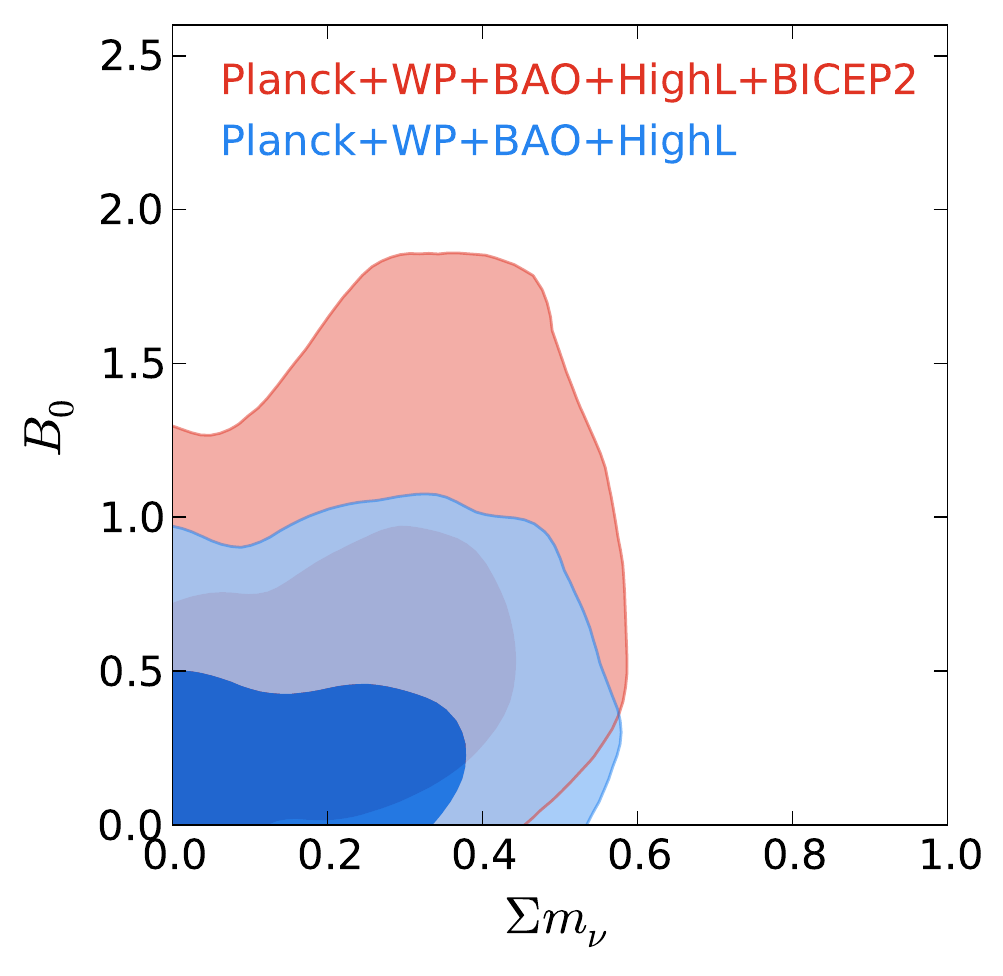}
\caption{Marginalized two-dimensional likelihood ($1, 2\sigma$ contours) constraints on $B_0$ and $\sum m_{\nu}$. It is clear that the degeneracy between $B_0$ and $\sum m_{\nu}$ sharpens if $B_0\gtrsim 1 $ for the data sets of Planck+WP+BAO combination. When adding the BICEP2 data, apparently it does not show significant improvement for the constraints on $B_0$ and $\sum m_{\nu}$. }\label{figone}
\end{figure}

\begin{figure*}
\includegraphics[width=5in,height=2.8in]{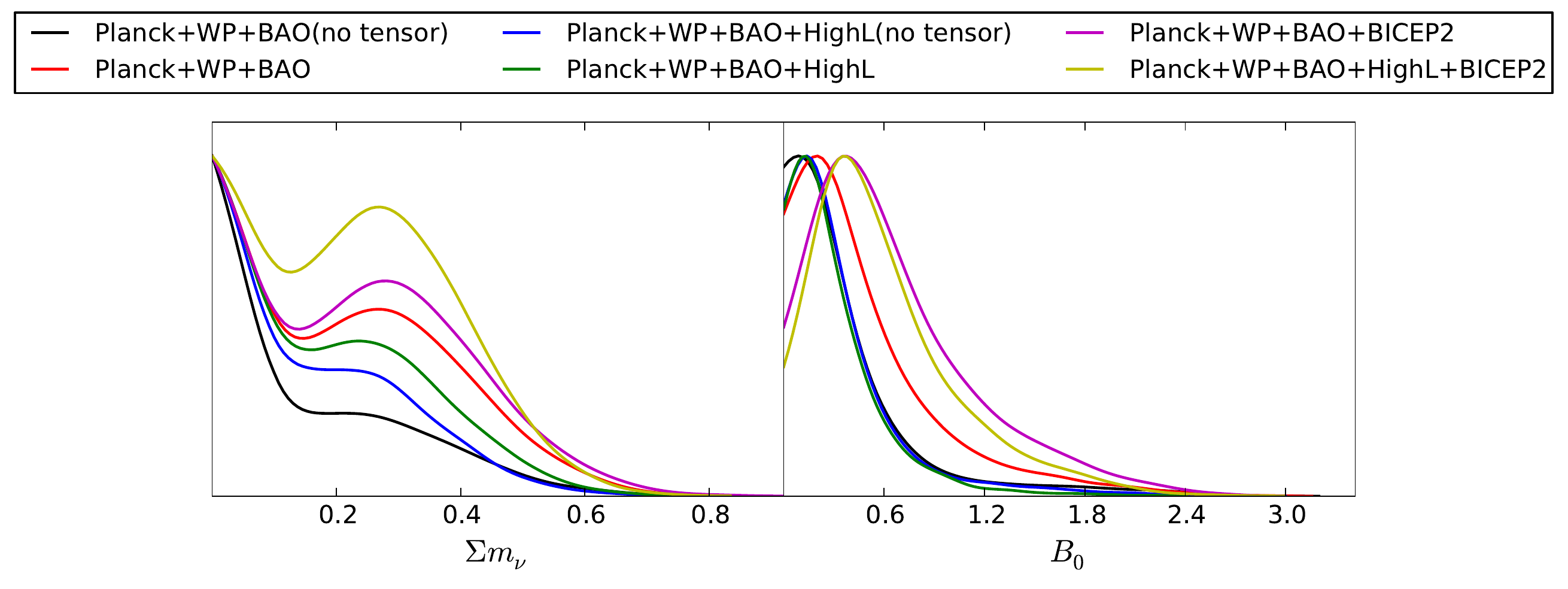}
\caption{One-dimensional marginalized likelihood on $\sum m_{\nu}$ and $B_0$. It is clear that the BICEP2 data tends to favour a non-zero value for the neutrino mass at around $\sum m_{\nu}\sim 0.3 {\rm eV}$ and a non-zero best-fitted value for $B_0$. }\label{oned}
\end{figure*}

In fact, the BICEP2 data does break the degeneracy between the massive neutrinos and the $f(R)$ gravity. The BICEP2 data tends to favour a non-zero value for the massive neutrino at around $\sim 0.3 {\rm eV}$ in the $f(R)$ gravity despite this feature does not show up in the fitting results for the $\Lambda$CDM model~\cite{xinzhang,hongli}. In the $\Lambda$CDM case, there is no strong degeneracy among massive neutrinos and other cosmological parameters as that in the $f(R)$ gravity. The massive neutrinos can be strongly constrained by the Planck and BAO data. The BICEP2 data does not significantly affect other cosmological parameters except for the tensor to scalar ratio $r$~\cite{xinzhang}.
The effect of the BICEP2 data only becomes manifest when the data sets such as the Planck and BAO data fail to break the degeneracy associated with the massive neutrinos(e.g. the $f(R)$ gravity case). In order to show this point, in Fig.~\ref{oned} we present the one-dimensional marginalized likelihood for the parameters $\sum m_{\nu}$ and $B_0$. In the left panel of Fig.~\ref{oned}, it is evident that the probability functions for the constraints with the BICEP2 data on $\sum m_{\nu}$ have significant secondary peaks, which indicates that the BICEP2 data tends to favour a non-zero value for the neutrino mass at around $\sum m_{\nu}\sim 0.3 {\rm eV}$ in the $f(R)$ gravity. Meanwhile, a non-zero value for the Compton wavelengths $B_0$ is also favoured at $1\sigma$ confidence level by the BICEP2 data, as shown in the right panel of Fig.~\ref{oned}. The best-fitted values for $B_0$ at the $68\%$ confidence level are
\begin{eqnarray}
B_{0} &=& 0.68^{+0.15}_{-0.63} \quad(68\%{\rm C.L.};{\rm Planck+WP+BAO+BICEP2})\quad,\nonumber\\
B_{0} &=& 0.63^{+0.15}_{-0.55} \quad(68\%{\rm C.L.};{\rm Planck+WP+BAO+HighL+BICEP2})\quad.\nonumber
\end{eqnarray}
However, the degeneracy between $\sum m_{\nu}$  and $B_0$ is not fully broken at the $95\%$ confidence level due to the insufficient accuracy of the measurement on the B-modes at the moment.
\begin{figure}
\includegraphics[width=3in,height=2.5in]{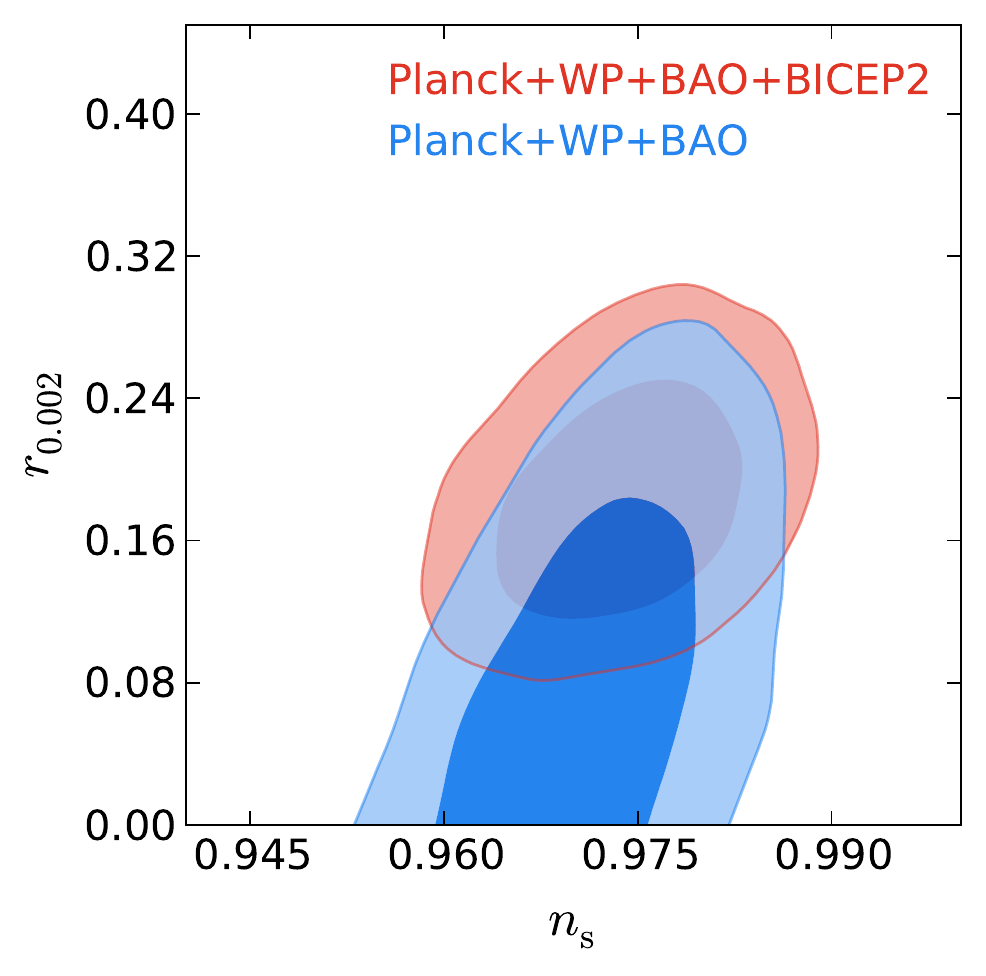}
\includegraphics[width=3in,height=2.5in]{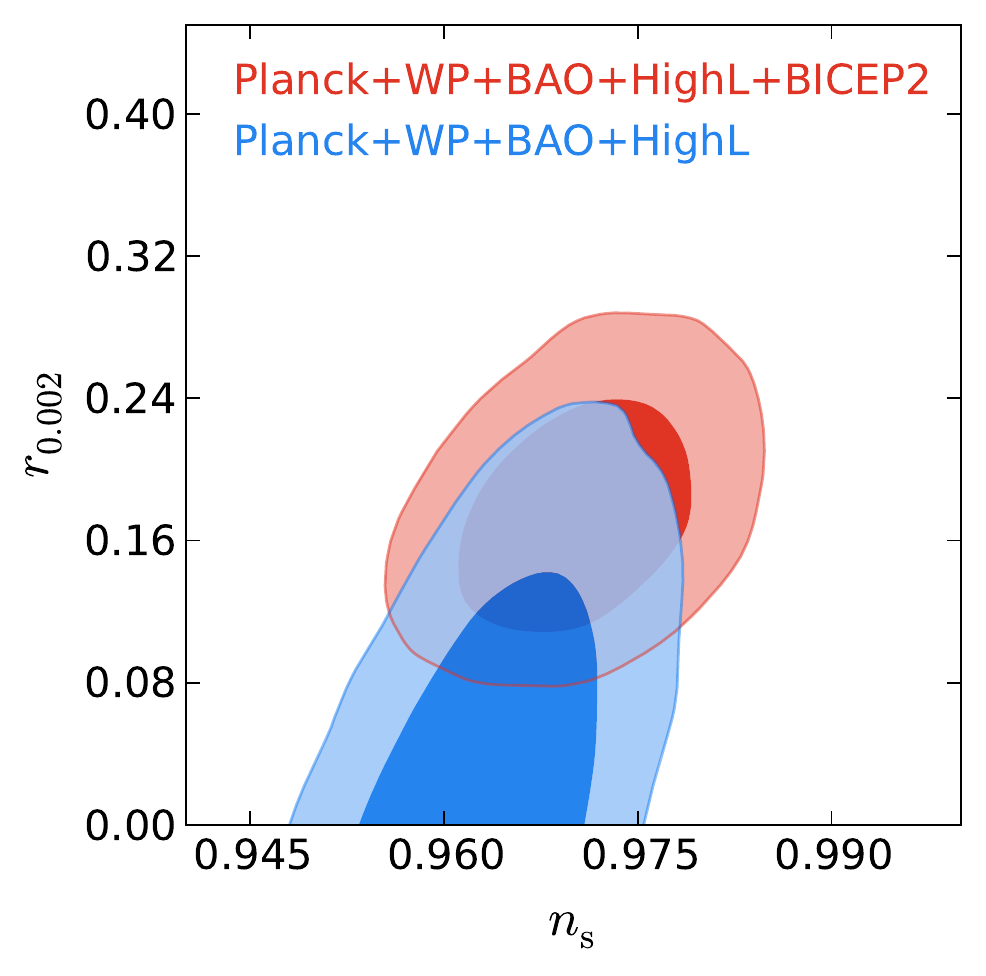}
\caption{Marginalized two-dimensional likelihood ($1, 2\sigma$ contours) constraints on $r_{0.002}$ and $n_s$. The 1$\sigma$ range of contours overlap with each other and the tension between Plank and BICEP2 is significantly reconciled. }\label{tension}
\end{figure}

Finally, we shall examine whether the $f(R)$ model, in the presence of massive neutrinos, can help to reconcile the tension on the tensor-to-scalar ratios between the measured values from Plank and BICEP2 as reported in the literature~\cite{xinzhang,hongli}. The fitting results on the ratio $r_{0.002}$ at the pivot scale $k_{s0}=0.002{\rm Mpc^{-1}}$ with the $68\%$ confidence level are
\begin{eqnarray}
r_{0.002} &<& 0.121 \quad(68\%{\rm C.L.};{\rm Planck+WP+BAO})\quad,\nonumber\\
r_{0.002} &<& 0.091 \quad(68\%{\rm C.L.};{\rm Planck+WP+HighL+BAO})\quad,\nonumber\\
r_{0.002} &=& 0.183^{+0.038}_{-0.040}  \quad(68\%{\rm C.L.};{\rm Planck+WP+BAO+BICEP2})\quad,\nonumber \\
r_{0.002} &=& 0.176^{+0.037}_{-0.048}  \quad(68\%{\rm C.L.};{\rm Planck+WP+BAO+HighL+BICEP2})\quad.
\end{eqnarray}
In Fig.~\ref{tension}, we show the marginalized two-dimensional likelihood ($1, 2\sigma$ contours) constraints on $r_{0.002}$ and $n_s$.
It is very interesting to find that the tension on the tensor-to-scalar ratio between Plank and BICEP2 is significantly reconciled in our model.
The 1$\sigma$ range of the contours from Planck and BICEP2 overlap with each other in both cases with and without including the HighL data sets.
\section{Conclusions\label{con}}
In this study, we have constrained the neutrino mass in the $f(R)$ gravity by using the latest observations from the Planck, BAO and BICEP2 data. We find that the measurement on the B-modes is a promising way to break the degeneracy between the massive neutrinos and the $f(R)$ gravity in the linear growth history. The tension on the tensor-to-scalar ratios between the measured values from Plank and BICEP2 is significantly reconciled in our model. We also find a large and non-zero value of the Compton wavelengths $B_0$ at $1\sigma$ confidence level for the $f(R)$ model in the presence of massive neutrinos when the BICEP2 data is used. However, this large value of the Compton wavelengths $B_0$ could be in tension with the stringent local\cite{HuI} and astronomical\cite{Jain} tests of gravity despite the current analysis of these tests does not take into account the effect of massive neutrinos. To evaluate the effect of massive neutrinos in the local tests of gravity is an object of our future work.

\emph{Acknowledgment: J.H.He acknowledges the financial support from the Italian Space Agency (ASI), through contract agreement I/023/12/0.
 }

\end{document}